\documentclass[preprint,proceedings]{rmaa}

\suppressfulladdresses % by popular request
\usepackage{paralist}
\usepackage{psfrag,color}

\SetYear{2006}
\SetConfTitle{First Light with the GTC}

\title{Star Formation Near Supermassive Black Holes} 

\author{
  Jonathan C. Tan\altaffilmark{1}}

\altaffiltext{1}{Department of Astronomy, University of Florida, Gainesville, FL 32611, USA.}

\shortauthor{Tan}
\shorttitle{Star Formation Near Supermassive Black Holes}

\listofauthors{Tan}
\indexauthor{Tan, J. C.}

\abstract{Supermassive black hole accretion and star formation appear
intimately connected. I review the observational and theoretical
evidence for this statement. I then discuss how focussed studies of
two systems, our Galactic Center and the nucleus of M87, can help to
improve our understanding of these processes.}

\resumen{La acreci\'on de los agujeros negros supermasivos y la
formaci\'on estelar parecen estar \'intimamente
relacionados. Eval\'uo la evidencia te\'orica y
observable de esta afirmaci\'on y expongo c\'omo
estudios espec\'ificos de dos sistemas, nuestro Centro
Gal\'actico y el n\'ucleo de M87, pueden ayudarnos a
comprender mejor estos procesos.}

\addkeyword{accretion --- galaxies: active --- stars: formation}

\begin{document}
\maketitle

\section{Observational Evidence}

Young, massive stars are seen within $0.5\:{\rm pc}$ of Sgr~A$^*$
in our Galactic center (Lu et al. 2005; Eisenhauer et al. 2005). How
they formed there is considered in \S\ref{S:GC}. An A-star spectrum,
indicating relatively young stars, is observed in M31's nucleus
(Bender et al. 2005). Late-type disk galaxies often have massive
nuclear star clusters (Walcher et al. 2005), although it is usually
difficult to tell if supermassive black holes are also present. In
earlier-type galaxies the black hole mass is $\sim$0.6\% of the
spheroid stellar mass (Magorrian et al. 1998), indicating that star
formation has proceeded in step with black hole growth. When we see
black holes growing, i.e. as quasars, there is often an IR bump in
their spectra (Elvis et al. 1994), which could be due to star
formation (Sirko \& Goodman 2003). Quasar spectra also show that gas
near the black hole has high metallicity (Hamann et al. 2002),
presumably because of massive star formation. Star formation is
observed in the nuclei of Seyfert galaxies (Terlevich 1996). In
luminous IR galaxies (LIRGs) large amounts of gas are driven into the
nuclei and star formation produces much of the luminosity
(Alonso-Herrero et al. 2006). Ultra-luminous IR galaxies (ULIRGs) are
more extreme examples of such systems, containing massive
circumnuclear disks of molecular gas thought to be self-regulated
against gravitational collapse by star formation (Downes \& Solomon
1998). Radio supernovae have been observed in the nuclei of ULIRG Arp~220,
occurring at a rate of at least 4 per year (Lonsdale et al. 2006).

\section{Theoretical Expectation}

If a gas disk has a high enough surface density, $\Sigma$, and is
sufficiently cool (i.e. if the effective sound speed, $c_s$, is small
enough), then it will be gravitationally unstable.
The Toomre parameter describing this condition for a Keplerian disk is
\begin{equation}
\label{Q}
Q \equiv \frac{c_s \Omega}{\pi G \Sigma} = \frac{3\alpha c_s^3}{G\dot{M}} =
 0.71 \alpha \frac{c_{s, 10\:{\rm km\:s^{-1}}}^3}{\dot{M}_{M_\odot\:{\rm yr^{-1}}}},
\end{equation}
where instability occurs for $Q<1$. Here we substituted $\Sigma$ for
the mass accretion rate, $\dot{M}=3\pi \nu \Sigma$ of a disk with
viscosity $\nu = \alpha c_s h$ and scaleheight $h$ (Goodman 2003). The
Shakura-Sunyaev viscosity parameter $\alpha$ takes values of about
unity in self-gravitating disks. Gammie (2001) finds $\alpha\simeq
0.3$ from 2D simulations of local patches of a disk. Global
simulations of disks around massive protostars indicate that the
effective viscosity due to spiral density waves is about an
order of magnitude larger than this (M. Krumholz, priv. comm.). If the
over-dense regions created by gravitational instability have a high
enough cooling rate, they should fragment into
stars.

The surface temperature of an optically thick disk heated solely by
viscous dissipation scales as $r^{-3/4}$ and the central temperature
falls off at a similar or even faster rate depending on the energy
transport mechanisms operating in the disk and the opacity of the
gas. For example for constant opacity and radiative diffusion, the
midplane temperature falls as $r^{-9/10}$ and so, for thermal pressure
support, the effective sound speed scales as $c_s\propto
r^{-9/20}$. For a typical quasar with $M=10^8\:M_\odot$ accreting at
the Eddington rate, the radius at which $Q$ drops to unity is $\sim
10^{-2}\:{\rm pc}$ (Goodman 2003). 
%(ref to Duellomond???) 
Viscous dissipation can only stabilize the very inner regions of these
accretion disks. In the scenario we consider to be most likely for
Galactic center star formation (\S\ref{S:GC}), viscous dissipation is
relatively more important since the young stars are forming very close
to the black hole.
%, though Levin (2006) showed it was still
%unlikely to be the dominant heating mechanism more than 0.06~pc from
%the black hole. 
In the sub-Eddington, $\sim 100$~pc disk of M87 (\S\ref{S:m87}),
heating due to viscous dissipation is much too small to stabilize the
disk (Tan \& Blackman 2005).

A number of additional factors may enhance disk stability. The outer
disk may be heated by radiation liberated in the inner disk close to
the black hole, although in practice this is not a very significant effect
(Goodman 2003).
%(see also the treatment of Kratter \& Matzner (2006) in
%the context of massive protostellar disks). 
The basic problem is that little of the flux from the central source
will be intercepted by a geometrically thin disk. If the equatorial
region and disk are also optically thick, then the flux will tend to be
beamed out in the polar directions. 

The disk will also be heated by the general radiation field present at
the center of its galaxy. Levin (2006) considered this for our
Galactic center, using a luminosity from young stars in the inner
$\sim {\rm pc}$ of $\sim10^7\:L_\odot$ and concluding the disk would
be heated to about 80~K, which dominates viscous heating beyond
$\sim$0.06~pc. However, the luminosity from young stars is expected to show
temporal fluctuations depending on the recent star formation activity.
In the center of an elliptical galaxy the main sources of irradiation
are the older stellar population and x-rays from the hot interstellar
medium (ISM).

In the local Galactic ISM, thermal pressure is not the dominant
component of support and magnetic fields, with pressure $P_B \propto
B^2$, play a significant role (Heiles \& Crutcher 2005). This may also
be true in circumnuclear disks. If magnetic flux, $B$, is frozen into
the gas, then $P_B \propto \Sigma^2$, which is enough to compensate
for the increasing weight of the gas $P_G \simeq G\Sigma^2$ in a
compressed region. For fragmentation and star formation to be able to
occur in a disk in which magnetic fields are important for stability
on large scales, we expect that magnetic flux must diffuse out of the
over-dense regions. If the field cannot diffuse significantly, this is
equivalent to a small effective cooling rate for the gas.  Magnetic
field will tend to diffuse out of dense gas, either by ambipolar or
turbulent diffusion. We expect that relatively strong magnetic
fields will act to slow the rate of star
formation in a globally unstable disk, but will be unlikely to prevent
it completely.

Thus, the theoretical expectation is that circumnuclear disks are
gravitationally unstable, except in their very central regions, and so
form stars.  Star formation helps to stabilize a disk by both consuming
gas, which reduces $\Sigma$, and heating the remainder, which raises
$c_s$.  Tan \& Blackman (2005) considered these two effects in the
context of Bondi-fed accretion disks around the supermassive black
holes in giant elliptical galaxies. Thompson, Quataert \& Norman
(2005) modeled the stability of disks with higher accretion rates,
applicable to quasar and ULIRG systems. In both cases, for disks that
self-regulate to $Q=1$, most of the mass flux being fed to the disk at
its outer edge turns into stars.

The above models must assume a heating rate per mass of stars formed,
which depends on their initial mass function (IMF). Conditions in the outer
parts of circumnuclear disks are not expected to be that different
from typical Galactic star-forming regions, with the exception of
somewhat higher ambient temperatures and pressures. The Bonnor-Ebert
mass, which is the maximum mass of a stable isothermal sphere of gas
and may be related to the peak of the stellar IMF, scales as $T^2
P^{-1/2}$, so the higher temperatures and pressures of circumnuclear
disks tend to counteract each other in this regard.

The disk orbital time scales as $t_{\rm orbit} \propto r^{3/2}$ while
the local star formation time for a fixed stellar mass (McKee \& Tan
2003) scales as $t_{*f} \propto P^{-3/8} \propto \Sigma^{-3/4}\propto
r^{9/8}$, where the last proportionality assumes the structure of a
constant $Q$ disk with viscosity proportional to total pressure
(Goodman 2003). We expect $t_{*f}\ll t_{\rm orbit}$ in the outer disk,
but these may become comparable to each other in the inner regions. In
this case, the collapsing fragment is likely to gain significant mass
by sweeping up gas from the disk in a runaway process only halted once
a gap is opened up. In quasar accretion disks this process may lead to
supermassive stars (Goodman \& Tan 2004), while in the Galactic center
(Nayakshin 2006) it may help to explain the IMF, which is observed to
be top-heavy (Nayakshin et al. 2006).

\section{The Galactic Center}\label{S:GC}

The distribution of many of the young stars near Sgr~A$^*$ in a
disk-like system and the lack of a larger-scale distribution suggests
formation {\it in situ}, rather than inspiral as part of a cluster.
Star formation in this region does not appear to be active today, so
we can view this process as episodic, with the last burst occurring
about 5-10~Myr ago. Dense, probably bound, molecular gas clumps are
observed within a few pc of Sgr~A$^*$ (Herrnstein \& Ho 2005;
Christopher et al. 2005).  The scenario we envisage to explain the
recent Galactic center star formation is the infall of such a clump
after having its orbit perturbed by either a cloud collision or
interaction with a supernova explosion. As the clump approaches the
black hole, it is tidally disrupted. Some of the gas is shocked and
then cools to form a disk or ring.
%, while other gas is able to continue in a stream on an eccentric orbit. The gas in this stream may accrete into the disk on a later orbital passage. 

As an example of this process, consider a clump with $2\times
10^4\:M_\odot$ that is able to feed half its mass into a disk with
radial size of 1~pc, so that $\bar{\Sigma}=0.66\:{\rm
g\:cm^{-2}}$. Assume the gas disk can cool to an effective
temperature of $100$~K (lower temperatures are hard to achieve given
the local stellar radiation field) with effective sound speed of
0.6~$\rm km\:s^{-1}$. Evaluating the Toomre parameter we find
\begin{equation}
\label{Q2}
Q \equiv 2.0 c_{s, {\rm km\:s^{-1}}} \frac{M_{\rm BH, 4\times 10^6\:M_\odot}^{1/2}}{r_{\rm pc}^{3/2} \Sigma_{\rm g\:cm^{-2}}},
\end{equation}
and the disk would be close to the threshold of instability.  However,
this simple analysis does not allow for the fact that, if the
viscosity is relatively low, the gas will tend to pile up at the outer
edge of the disk, forming a ring with a much higher surface
density. Numerical simulations, that adequately resolve the formation
of a circumnuclear disk and the star formation process within it, are
required in order to investigate this scenario in more detail.

\section{Bondi-Fed Disks in Ellipticals: M87}\label{S:m87}

The black holes in the centers of giant elliptical galaxies should
accrete interstellar gas by Bondi accretion, and this expected
accretion rate can be inferred from x-ray observations of the density
and temperature of the gas in nearby systems like M87 (Di~Matteo et
al. 2003). With this constraint on the mass supply rate to the
disk, such systems present special opportunities for testing models of
disk accretion, possible star formation and black hole fueling. 

Tan \& Blackman (2005) considered the gravitational stability of the
accretion disk in M87, which is $\sim$100~pc in radial extent and thought
to be fed at $\sim0.04-0.15\:M_\odot\:{\rm yr}^{-1}$. They predicted the
disk should be unstable and form stars. A simple model for
self-regulation ($Q\simeq 1$) of the disk by star formation led to the
prediction of a cold molecular gas mass of $1-5\times 10^6\:M_\odot$.
Observations of CO(J=2-1) with the Sub-MM Array yield mass
estimates of $4.3\pm1.2 \times 10^6\:M_\odot$ in agreement with this
prediction (Tan et al. 2006), although higher sensitivity data
are required for definitive confirmation.

Additional predictions of the model are thermal emission from dust in
this gas and emission from young stars forming at $\sim0.01 -
0.1\:M_\odot\:{\rm yr^{-1}}$. Perlman, Mason, Packham et al. (2007,
these proceedings) reported a 60~K thermal component in the nucleus
with luminosity $4\times 10^{38}\:{\rm ergs\:s^{-1}}$. This is much
smaller (by factors of $\sim 10^3$) than the luminosity expected from
the above star formation rates, but more luminosity may be being
released from cooler dust that is not probed by the 35~$\rm \mu m$
longest wavelength of the Perlman et al. observations. The H$\alpha$
luminosity of the M87 disk is consistent with the above star formation
rate (Tan \& Blackman 2005), although it is difficult to assess how much of this is
due to ionization by the AGN (Kim, Ho \& Im 2006).

The M87 nucleus is underluminous by a factor of $\sim$100 compared
to models of thin disk accretion at the Bondi rate. It
is possible that star formation, by acting as a mass sink, helps to
explain this discrepancy (Tan \& Blackman 2005). It is also possible
that in the central regions of the disk there is a transition to a
radiatively inefficient accretion flow.

\end{document}